\documentclass{ws-procs9x6}

\begin{document}

\title{Casimir forces between spheres and loop integrals}

\author{James Babington$^*$}

\address{Quantum Optics and Laser Science, Blackett Laboratory, Imperial College London,\\
Prince Consort Road, London SW7 2AZ, U.K.\\
$^*$E-mail: j.babington@imperial.ac.uk}



\begin{abstract}
A summary of recent calculations of Casimir forces between a collection of $N$-dielectric spheres is presented. This is done by evaluating directly the force on a sphere constructed from a stress tensor, rather than an interaction energy. A loop integral formulation is also discussed where we rewrite the expressions for the force in terms of loop integrals for the effective classical propagation of the electric and magnetic fields.
\end{abstract}


\bodymatter

\section{Introduction}

An important step in better understanding Casimir forces and their application to nano-scale environments is the nature and influence of the geometries used. If we consider a collection of small bodies and their interactions, a quantitative understanding is necessary to address any issue of applications. One can calculate the forces between bodies directly by using a stress tensor evaluated on the body in question. This is an experimentally accessible prediction of the theory e.g. three body forces can be measured between dielectric spheres~\cite{PhysRevLett.92.078301} for critical Casimir forces. The choice of stress tensor in media is not unique and depends on what consistency criteria is used. Typically the differences will show up both in the scale of the forces and the higher order curvature corrections, and one may view this as a way of finding the correct low energy description.

To address the nature of such interactions between bodies~\cite{emig2008, Rahi:2009hm,kenneth:014103,Bulgac:2005ku} Casimir interaction energies have been evaluated for collections of compact objects interacting with different force carrying fields (electromagnetic and scalar). The approach taken has been to evaluate a suitable energy functional integral using a T-matrix, whereby an interaction energy can be deduced, normalised with respect to their energy when separated at infinity.

In this talk I summarise recent work we have done on calculating the Casimir force on a single sphere, in an $N$-sphere setup~\cite{James2009}. By using a multiple scattering approach to evaluate essentially the  classical scattering Green's function of the configuration, we are able to evaluate the force directly on the sphere. The total path length plays a fundamental role and what leads one to a loop description~\cite{Jamesloop2009}. Similar loop descriptions have appeared before in~\cite{PhysRevA.58.935,schaden:042102,Scardicchio2005552} where a semi-classical type expression of the density of states is found, and in~\cite{Gies:2001tj} a worldline description for scalar fields is given.

\section{The $N$-Sphere configuration}

The question we are addressing is how to calculate the force on a particular sphere as a result of all the interactions with the remaining spheres in a particular static configuration. The Casimir force on a sphere (given by the ball $B^2$) in the $j$-direction due to the effects of the $N$-sphere system is given by
\begin{equation}
\label{eq:stressintegral}
F^j(1|N-1)=\int_{B^2} d^3x \nabla_i T^{ij}(x).
\end{equation}
The stress tensor we choose is the standard vacuum expression (which is consistent with the Lorentz force law~\cite{raabe:013814})
\begin{equation}
T_{ij}(x)=\mathbf{E}_i(x)\mathbf{E}_j(x)+\mathbf{B}_i(x)\mathbf{B}_j(x)-\frac{1}{2}\delta_{ij}(|\mathbf{E}(x)|^2+|\mathbf{B}(x)|^2),
\end{equation}
where  $x \in B^2$ and it is understood that we are taking the limit for the initial and final points. We then need to evaluate the scattering correlation functions (whilst dropping the direct modes of propagation)
\begin{equation}
\lim_{y \rightarrow x}\mathbf{E}_i(x)\mathbf{E}_j(y)=\int_{0}^{\infty}\int_{0}^{\infty}d\omega d\omega^{\prime}\langle \mathbf{E}^{out}_{i}(x;\omega)^{\dagger}\mathbf{E}^{in}_{j}(y;\omega^{\prime})\rangle,
\end{equation}
and similarly for magnetic fields. To construct the scattering two point function we write the fields in a mode decomposition~\cite{Mackowski06081991} of spherical vector wave functions that are centred on each sphere centre. Then by applying the standard continuity equations at each of the spheres surfaces, one can calculate the out modes in terms of the in modes and scattering (Mie) coefficients. Assuming that the background in which we are evaluating this is filled with quantum noise such that the noise-current two point function is non-zero we find for the $N$-body force on a sphere (suppressing the $SO(3)$ indices)
\begin{eqnarray}
\label{eq:NSPHEREFORCE2}
\mathbf{F}[1|N-1]&=&-(-1)^N\frac{\hbar }{4\pi } R[1]
 \Im \int_{0}^{\infty}d\omega k \coth (\hbar \omega /K_BT)\langle\mathbf{1}| [ \alpha^{1}(\omega R[1]) \nonumber \\ 
& &\times 
\sum_{i=2}^{N}A^{1,i}(\mathbf{r}[1,i]) \cdot \alpha^{i}( \omega R[i] )\cdots 
\cdots \sum_{j=2}^{N}A^{i,j}(\mathbf{r}[i,j]) \cdot \alpha^{j} \nonumber \\
&&\times \sum^N_{j=2}\nabla_{\mathbf{r}[j,1]} A^{j,1}(\mathbf{r}[j,1])] j(kR[1])
h^{+}(kR[1])W(\omega R[1]) |\mathbf{1}\rangle. \nonumber \\
&\equiv&  -(-1)^N\frac{\hbar }{4\pi }\sum^{N}_{i=2} \nabla^{last}_{\mathbf{r}[i,1]}\int_{0}^{\infty}d\Omega  \cot \left(\frac{\hbar \Omega}{k_BT}\right) \mathcal{Z}[\alpha,A,W].
\end{eqnarray}
Here,  $\alpha^i(\omega R[i])$ are the Mie scattering coefficients in the $SO(3)$ basis for sphere $i$ with radius $R[i]$; $A^{i,j}(\mathbf{r}[i,j])$ are the translation matrices mapping the TE and TM vector wave functions between spheres $i$ and $j$; the vectors $| \mathbf{1}\rangle$ give the truncation in the $L$ angular momentum quantum number (leading to a multipole type expansion); and the $W$ is just the collection of the four different contributions that make up the stress tensor, together with the two necessary Bessel functions evaluated on the spheres surface. In the last line a Wick rotation to imaginary frequencies has been performed and the $\mathcal{Z}$-function has been defined for later reference.

Note the explicit form of the translation matrices involve exponentials of the inter-sphere separations~\cite{James2009} and thus it is the \emph{total path length} that plays the key role in understanding the variables of the system. For simple setups (e.g. two and three sphere systems) we can evaluate Equation~(\ref{eq:NSPHEREFORCE2}) in different perturbative regimes e.g. retarded or non-retarded limits using static values for the permittivities (see~\cite{James2009} for explicit evaluations and force plots of two and three sphere systems).

\section{Loop integrals}

Following on from the observation that Equation~(\ref{eq:NSPHEREFORCE2}) involves the total path length in the form of a loop, in~\cite{Jamesloop2009} an attempt is made to develop this further. It is similar to the path integrals used in~\cite{PhysRevA.58.935} and~\cite{Gies:2001tj} where a fictitious time is introduced as well as mass scale set equal to unity and an appearance of Planck's constant in the particle action. As pointed out in~\cite{Scardicchio2005552}, it is misleading to call this a semi-classical evaluation because of the absence of the dimensionful Planck's constant. Concurrent with this is the absence of a mass or length scale with which to define a dimensionless action.  

In~\cite{Jamesloop2009} a symbolic expression for the $\mathcal{Z}$-function was found that features the loop structure in an explicit manner
\begin{eqnarray}
\label{eq:Loop3}
\mathcal{Z}^{(s)}[\Omega,x]& = &\sum_{\mathcal{C}_x} \langle\mathbf{F}^s_x|  \exp \left[-\oint_{\mathcal{C}_x}dq^i\hat{D}_i(q,\Omega) \right] |\mathbf{I}^s_x\rangle \nonumber \\
&=& \sum_{\mathcal{C}_x}\mathbf{ Tr}\left(  \exp \left[-\hat{S}_R(\mathcal{C}_{x}, \Gamma, \Omega) \right] \right)_{\{\mathbf{I}^s_x,\mathbf{F}^s_x\}}
\end{eqnarray}
where $x$ is the initial and final point of the loop, and $|\mathbf{I}\rangle$ and $|\mathbf{F} \rangle$ are the initial and final states (i.e. boundary conditions imposed on the eigenfunctions used to represent in and out modes). The generator $\hat{D}_i$ of translations and the loop $\mathcal{C}_{x}$ implicitly depend on the background potentials. The connection $\Gamma$ is formed from the background potentials in which the field propagates. One now needs to find a representation of this object. In fact it can be given a path integral representation, albeit a classical one i.e. no $\hbar$ featuring anywhere. One first needs to invert the Helmholtz operator (here partial derivatives have been promoted to covariant derivatives w.r.t. the permittivity and permeability)
\begin{eqnarray}
\label{eq:twopointf}
\mathbf{\Delta}(z,y)&=&\langle z |\left[-c^2/\Omega^2 \nabla_{\epsilon}\wedge\nabla_\mu \wedge -\mathbf{1} \cdot \right]^{-1}| y \rangle \nonumber \\
&=& \int_0^{\infty}d\tau \langle z |e^{-\tau [ c^2/\Omega^2\nabla_{\epsilon}\wedge\nabla_\mu\wedge +\mathbf{1}\cdot]}| y \rangle.
\end{eqnarray}
Introducing a world line metric $e$ to implement the Helmholtz equation on physical states, and an integration over the $\hat{\nabla}_i$ operators (again acting on physical states), together with an integration over paths one finds for the scalar version of the Helmholtz operator (i.e. two potentials but no spatial indices)
\begin{eqnarray}
 \mathcal{Z}[\Omega,x] =\int^{1}_{0} d\tau \sum^{\infty}_{n=1} \langle\mathbf{F}_x | \oint_{q[0]=q[1]=x} [dq]\sqrt{\det [\Omega^2/c^2(\epsilon \cdot \mu)]} \nonumber \\
\cdot \int [d\hat{\nabla}] [de]\exp \left(-n \int^{\tau}_{0}dt[ \dot{q} \cdot \hat{\nabla}+e(\hat{\nabla}_{\epsilon}\cdot \hat{\nabla}_{\mu}+1)]\right) |\mathbf{I}_x\rangle,
\end{eqnarray}
and
\begin{equation}
\int^{\infty}_{0}d\Omega \mathcal{Z}[\Omega,x] \sim \sum_{loops} \mathcal{Z}[\Omega,x].
\end{equation}
If we perform first the integral over the world line metric, the Helmholtz equation is implemented. Performing the integral over the derivatives returns a configuration space path integral which would require gauge fixing the world-line metric. The classical equations of motion then lead to closed geodesics defining the loops. Integration over the frequency provides part of the sum over paths, whilst the tau integration gives the windings of the loops.
\section{Conclusions}
\label{sec:conclusions}
In this talk I have summarised recent work we have done on calculating Casimir forces between spheres using a multiple scattering approach. The total closed path length plays a key role in understanding the calculated forces and leads to a loop description. By considering the origin of the translation coefficients, together with a path integral representation of the Helmholtz operator, one is able to reformulate the loop integral as a sum over of all possible loops. 

\section*{Acknowledgments}
J.~B. wishes to thank Stefan Buhmann, Stefan Scheel, Alex Crosse, Rachele Fermani and John Gracey for numerous helpful and constructive discussions, and the organisers of QFEXT09 for the opportunity to present this material. This work was supported by the SCALA programme of the European commission.


\begin{thebibliography}{99}

\bibitem{PhysRevLett.92.078301}
M.~Brunner, J.~Dobnikar, H.-H.~von~Gr\"unberg, and C.~Bechinger,
Phys. Rev. Lett. \textbf{92}, 078301 (2004).

\bibitem{emig2008}
T.~Emig and R.~L.~Jaffe, J. Phys. A \textbf{41}, 164001 (2008).

\bibitem{Rahi:2009hm}
  S.~J.~Rahi, T.~Emig, N.~Graham, R.~L.~Jaffe, and M.~Kardar,
Phys. Rev. D \textbf{80}, 085021 (2009). 

\bibitem{kenneth:014103}
O.~Kenneth and I.~Klich, Phys. Rev. B \textbf{78}, 014103 (2008).

\bibitem{Bulgac:2005ku}
  A.~Bulgac, P.~Magierski and A.~Wirzba,
  Phys.\ Rev.\  D {\bf 73}  025007 (2006).

\bibitem{James2009}
J.~Babington and S.~Scheel",
arXiv:0909.3285 [quant-ph].

\bibitem{Jamesloop2009}
J.~Babington,
arXiv:0909.3315 [quant-ph].

\bibitem{PhysRevA.58.935}
M.~Schaden and L.~Spruch,
Phys. Rev. A \textbf{58}, 2, (1998).

\bibitem{schaden:042102}
M.~Schaden,
Phys. Rev A  \textbf{73}, 4, 042102, (2006).


\bibitem{Scardicchio2005552}
A. Scardicchio and R.L. Jaffe, 
Nucl. Phys B, \textbf{704}, 3, (2005).

\bibitem{Gies:2001tj}
H.~Gies and K.~Langfeld,
Int. J. Mod. Phys. \textbf{A17}, 966-978, (2002).

\bibitem{raabe:013814}
C.~Raabe and D.-G.~Welsch, Phys. Rev. A \textbf{71}, 013814 (2005).

\bibitem{Mackowski06081991}
D.~W.~Mackowski, Proc. R. Soc. Lond. A \textbf{433}, 599 (1991).  


\end{thebibliography}

\end{document}